\documentclass[12pt,letterpaper]{article}
\usepackage{graphicx}
\usepackage{graphics}
\usepackage{color}
\usepackage{cite}
\usepackage[cp1251]{inputenc}
\usepackage[english]{babel}
\usepackage[OT1]{fontenc}
\usepackage{amsmath}
\usepackage{amsfonts} 
\usepackage{setspace}
\usepackage{amsmath} 
\begin{document}

\title{New ellipsometric approach for  determining  small light ellipticities }
{  \author{ \small Nazar Al-wassiti1$^{2,3}$, Evelina Bibikova$^{1,2}$, Nataliya Kundikova$^{1,2,*}$\\
{\it\small $  ^1$Nonlinear Optics Laboratory, Institute of Electrophysics,} \\ {\it \small
Ural Branch of the Russian Academy of Sciences,} \\ {\small
\it 
 Ekaterinburg, 620016 Russia} \\
{\it \small $^2$Department of Optoinformation, South Ural State University,}\\ {\it \small
Chelyabinsk, 454080 Russia} \\
{\it \small $^3$Department of Physics,                                    
College of science,}\\
{\it \small Al-Mustansiriya University, Baghdad, Iraq} \\
{\small $^*$kundikovand@susu.ru}}
\date{}
\maketitle
}
\begin{abstract}
We propose a precise ellipsometric method for the investigation of coherent light with
a small ellipticity. The main feature of this method is the use of  compensators with  phase delays providing the maximum accuracy of measurements for the selected range of ellipticities and taking into account the interference of multiple reflections of coherent light. 
The relative error of the ellipticity measurement in the range  of $7.5\times 10^{ - 4} \div 4.5\times 10^{ -3}$ does not exceed $ 2\%$.

\end{abstract}


\section*{Introduction}
In general, any polarized light beam can be considered as a point at the Poincare sphere having the ellipticity and the ellipse azimuth. The ellipticity is the ratio of the minor to the major axis of the polarization ellipse traced by the electric field vector as a function of time, and varies from 0 for linearly polarized light to $\pm 1$ for circularly polarized light. The ellipse azimuth is the orientation of the major axis measured counterclockwise from the $x$ axis.
Linear polarized light can acquire ellipticity due to vacuum magnetic birefringence   \cite{ DellaValle2013, Zavattini2007 } due to strong interactions between chiral molecules and orbital angular momentum of light \cite{ Wu2015 } and due to reflection from an interferential mirror \cite{ Carusotto1989, Bielsa2009, Dupre2015}. Birefringence and dichroism can be induced in a tensile-strained bulk semiconductor optical amplifier \cite{ Guo2005 }.
The elliptical polarization of light appears in localized structures in broad area Vertical-Cavity Surface-Emitting Lasers \cite{ Averlant2016a, Averlant2016 }. Propagation through birefringent or chiral material,  reflection, or refraction at an interface causes linearly polarized light to also become elliptically polarized \cite{ Azzam2011 }. 
Ellipticity measurements provide information on the degree of alignment of optically anisotropic blocks \cite{ Kathrein2016 } and on birefringence of the primate retinal nerve fiber layer \cite{ RylanderIII2005}. 

Photometric methods comprising the Stokes parameters measurement and ellipsometric methods are used to determine the ellipticity of a light beam \cite{ Azzam2011, Guo2007, Shinki2012, Palma-Vargas2011}.
The values of the acquired ellipticity can be as small as $10^{-3}$ \cite{Averlant2016}. In principle, the known photometric methods make it possible to measure small ellipticities, but require precision photometric instruments and rather powerful lasers. Ellipsometric  methods require a compensator (retardation plate) and an analyser. If elliptically polarized light propagates through the compensator and the  analyser, light extinction can occur under proper orientation of the compensator and the analyser. To determine any state of polarization, a quarter wave plate with the phase retardation $\Gamma= \pi \left/ 2\right.$ is usually used as the compensator \cite{ Azzam2011, Azzam1987 }.

Here we present a new ellipsometric method suitable for small ellipticity measurements.
The main feature of this method is the use of  compensators with phase retardations providing the maximum accuracy of measurements for each selected range of ellipticities and taking into account the interference of multiple reflections of coherent light.
We have calculated the compensator phase retardation which provided the maximum method accuracy for all ellipticities and demonstrated how it is possible to avoid the influence of the interference of multiple reflections on ellipticity measurement results. We have measured ellipticities in the range of $7.5\times 10^{-4}\div  4.5\times 10^{-3}$ with a relative accuracy 
better  than $2\%$.


\section{Ellipsometric approach for determining  the polarization state of light with small ellipticity }
Let us describe the polarization state conversion using the Jones calculus  \cite{ Azzam1987, Gerrard75}. We will describe the polarization state of a fully polarized quasi-monochromatic light beam propagating along the $z$-direction by the Jones vector or the Maxwell vector:
\begin{equation*}
{\bf E}  =  \left( \begin{array}{c}
 E_{x}  \\ 
  E_{y}  \\  \end{array} \right)\exp\left(i \omega t\right).
\end{equation*}
Here, $i=\sqrt{-1}$, $E_{x}$ and $E_{y}$  are the complex components of the transverse electric vector along the $x$- and $y$- directions, respectively. The Jones vector for a linearly polarized beam with the angle $\beta$ ($ -\pi/4 \le \beta \le \pi/4 $)    between the direction of polarization and the $x$- direction has the form:
\begin{equation*}
{\bf E}_{\rm l}  =  \left( \begin{array}{c}
 \cos \beta  \\ 
  \sin \beta  \\  \end{array} \right).
\end{equation*}
The Jones vector for an elliptically polarized beam with the ellipticity angle  $ \varepsilon$  ($ -\pi/4 \le \varepsilon \le \pi/4 $) and  angle 
$ \varphi$ ($ -\pi/4 \le \varphi \le \pi/4 $)
between the major ellipse axis and the $x$-direction has the form:
\begin{equation}
{{\bf E}_{\rm e}} = \left( \begin{array}{c}
\cos \varepsilon   \cos \varphi  + i\sin\varepsilon \sin\varphi \\
 - \cos \varepsilon   \sin \varphi  + i\sin\varepsilon \cos \varphi 
\end{array} \right).
\label{ellips}
\end{equation} 	
	The ellipticity angle  $ \varepsilon$ is connected with light ellipticity
$e$  in the following way:
\begin{equation*}
e=b/a= \tan \varepsilon.
\end{equation*}
Here, $a$ and $b$ are the principal semi-axes of the polarization ellipse.
 
An optical system which linearly transforms the polarization state  is described by a $2\times 2$ Jones matrix with complex entries \cite{ Azzam1987, Gerrard75}. The Jones matrix of a linear phase anisotropy, or a birefringent plate with mutually orthogonal slow and fast axes (a compensator), is given by:
\begin{equation}
{\bf{\hat T}}^{\rm{C}} \left( {\Gamma} \right)= {\bf{\hat T}}^{\rm{LP}}\left( {\Gamma} \right)=\left( {\begin{array}{*{20}{c}}
1&0\\
0&{\exp (i\Gamma )}
\end{array}} \right). 
\label{phase}
\end{equation} 
The slow axis is directed along with the $x$-axis and  $\Gamma $ is the phase shift between two orthogonal linear components of the electric vector, $0 \le \Gamma  \le 2\pi $.
Taking into account the interference of multiple reflections, we get matrix 
${\bf \hat T}^{\rm{C}}\left( \Gamma,F \right)$  in the following form:
\begin{equation}
{\bf\hat T^{\rm{C}}}\left( {\Gamma},F \right)={\bf\hat T}^{\rm{LP}} \left( {\Gamma  } \right){\bf{\hat T}}^{\rm{LA}} \left( { F } \right)= \left( {\begin{array}{*{20}{c}}
1&0\\
0&{F\exp (i\Gamma )}
\end{array}} \right).
\label{phase2}
\end{equation} 
Here, ${\bf{\hat T}}^{\rm{LA}}$ is the Jones matrix of a linear amplitude anisotropy, $F$ is the relative absorption of two linear orthogonal components of the electric vector, 
$0\le F \le 1$ \cite{ PhysRevE.74.056607, Bibikova2013   }.

To determine the unknown values of $\varepsilon$ and  $\varphi$, let us transmit elliptically polarized light with the Maxwell vector described by Eq.(\ref{ellips}) through the compensator described by Eq. (\ref{phase2}). The  Maxwell vector  of the transmitted light beam is given by:
\begin{equation*}
 {{\bf E}_2} = {\bf{\hat T}}^{\rm{C}}\left( {\Gamma ,F} \right){{\bf E}_{\rm e}} =\left( \begin{matrix}
   \cos \varphi \cos \varepsilon +i\sin \varepsilon \sin \varphi   \\
   F\exp (i\Gamma )\left( -\sin \varphi \cos \varepsilon +i\sin \varepsilon \cos \varphi  \right)  \\
\end{matrix}\    \right)
\end{equation*}
with unknown values of $\varepsilon$ and  $\varphi$.

The light beam will be extinguished by an analyser if the beam is linearly polarized after emerging from the compensator. 
In this case we receive  the following equation:
\begin{equation}
{{\bf E}_{2}}  =\left( \begin{matrix}
   \cos \varphi \cos \varepsilon +i\,\sin \varepsilon \sin \varphi   \\
   F\exp (i\Gamma )\left( -\sin \varphi \cos \varepsilon +i\,\sin \varepsilon \cos \varphi  \right)  \\
\end{matrix}\    \right)=\left( \begin{matrix}
  \cos \beta  \\ 
  \sin \beta   
\end{matrix} \right).
\label{eq1}
\end{equation}
The solution of Eq. (\ref{eq1}) for the ellipticity angle $\varepsilon$ is given by two forms:
\begin{equation}
\sin 2\varepsilon =-\frac{2F\tan \beta \,\sin  \Gamma }{\left( {{F}^{2}}+{{\tan }^{2}}\beta  \right)}
\label{eq2}
\end{equation}
and
\begin{equation}
\tan 2\varepsilon =\sin 2\varphi  \tan \Gamma. 
\label{eq3}
\end{equation}
Equation (\ref{eq2}) allows us to determine the ellipticity angle $\varepsilon$ by measuring the angle $\beta$, but we should know the value of $F$. Usually Eq. (\ref{eq2}) is used in ellipsometric  methods.

We propose the use of Eq. (\ref{eq3}) for measuring ellipticity angle $\varepsilon$. The coefficient $F$ does not appear in Eq. (\ref{eq3}), which improves the accuracy of the measurements,
but requires measuring  the value of $\varphi$.

It follows from  Eq. (\ref{eq3})  that a polarized beam with the ellipticity $e = \tan \varepsilon $  will be extinguished under  two physically different positions of the compensator (the azimuths ${\psi _1}$ and ${\psi _2}$) related to any system of coordinates, and   $\left|\psi _1+\psi _2 \right|= 90^{\rm \circ }$. Consequently, to measure  the ellipticity  it will suffice to determine the null positions of the compensator in any system of coordinates and to  find the absolute value of $\varphi $:
\begin{equation}
\left| \varphi  \right| = \pi  \left/  4 \right.
  - \left| \psi _1-\psi _2 \right|\left/ 2 \right.
\label{eq05}
\end{equation}
and the absolute value of $\varepsilon $:
\begin{equation}
\tan \left(2\left|\varepsilon\right|\right) =\sin \left(2\left|\varphi\right|\right)  \tan \Gamma. 
\label{eq5}
\end{equation}
The sign of ellipticity $e = \tan \varepsilon $ is connected with the sign of $\beta$.

So, it is possible to determine light ellipticity $e=\tan \varepsilon$ by measuring angle $\left|\varphi \right|$ using a compensator with the known phase retardation $\Gamma $.

\begin{figure}[h]
\begin{center}
\includegraphics[scale=0.5]{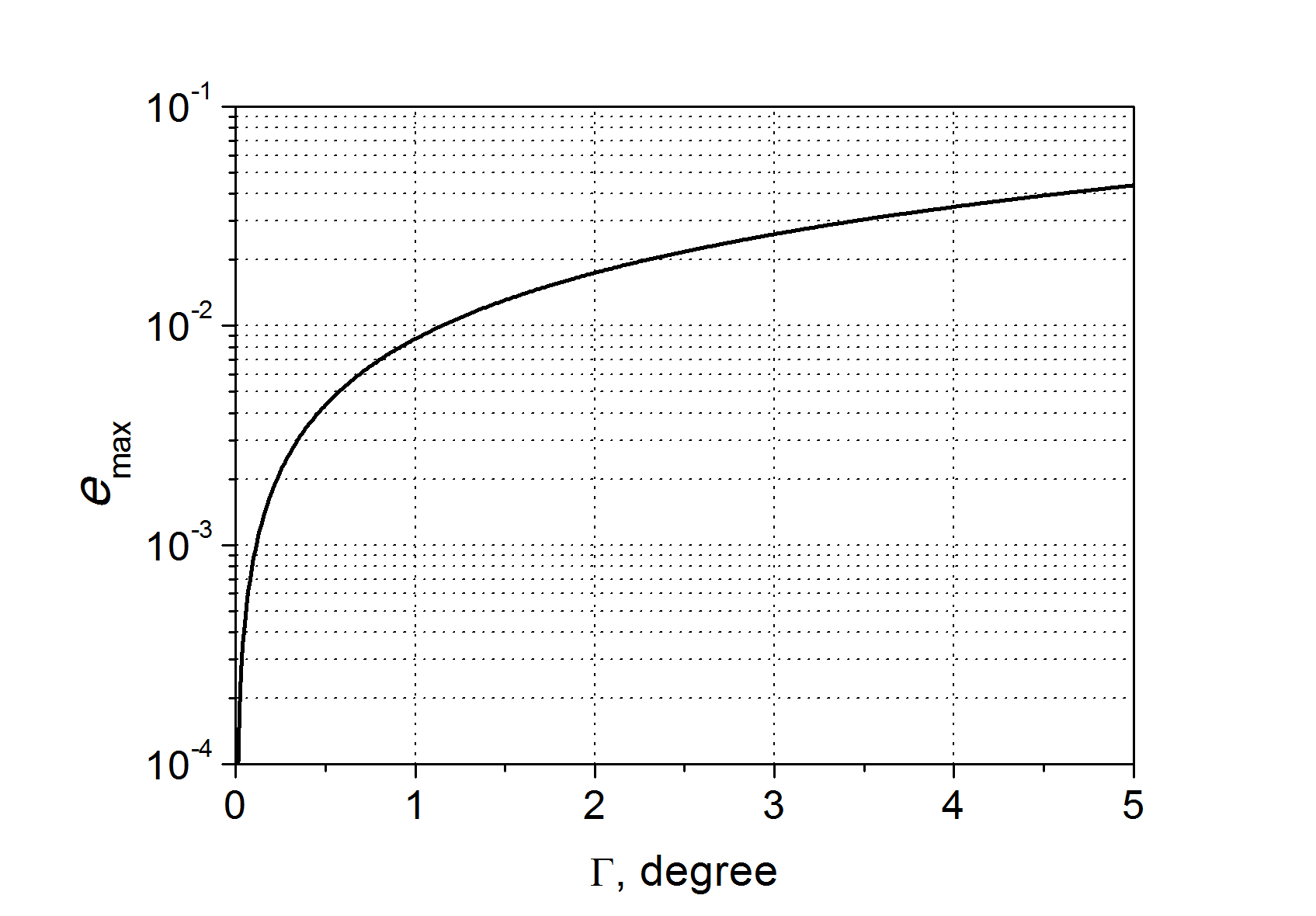}
\caption{ 
The dependences of the maximum measurable ellipticity $e_{{\rm{max}}}$ on the compensator phase retardation $\Gamma $.}
\label{pic2}
\end{center}
\end{figure}

To improve the accuracy of measurement within a certain range of ellipticities we should determine the proper phase retardation of the compensator. The compensator phase retardation $\Gamma $ should be reasonably large to compensate  light ellipticity during the process of the measurement. It follows from Eq. \ref{eq3} that the measured ellipticity is limited from above:	
\begin{equation}
{e}_{\max } = \tan\left(\Gamma \left/  2 \right.\right).
\end{equation}
The dependence  of the maximum measurable ellipticity $e_{{\rm{max}}}$ on the compensator phase retardation  $\Gamma $ is shown  in Fig. \ref{pic2}.

\begin{figure}[h]
\begin{center}
\includegraphics[scale=0.5]{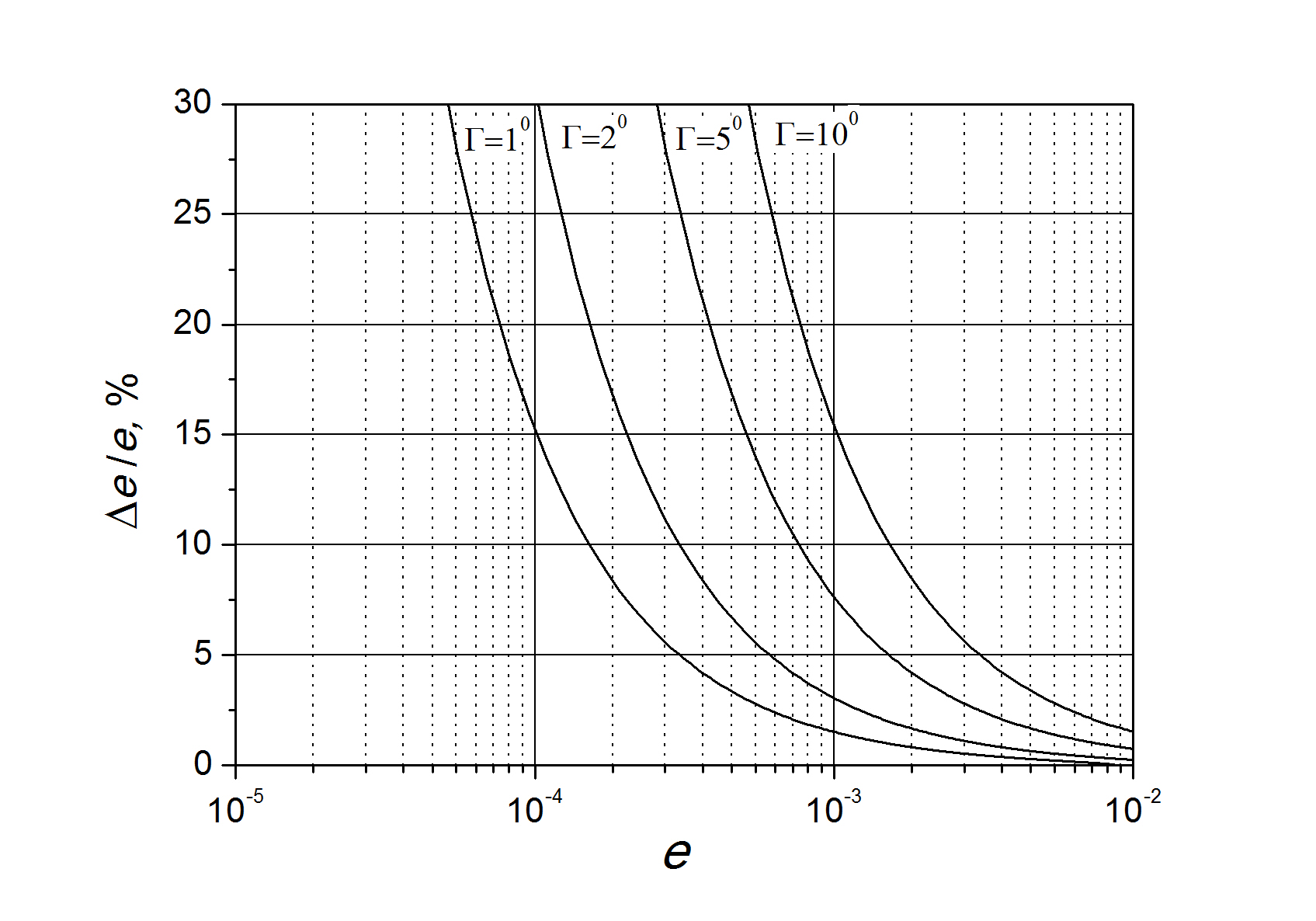}
\caption{ Dependence of the value 
$\left|   \Delta e\left/e \right. \right|$ 
 on  the ellipticity ${e}$    under measurements with compensators of different phase retardations $\Gamma $.}
\label{pic1}
\end{center}
\end{figure}

Figure \ref{pic1}
shows  the calculated dependences of the relative error of the ellipticity measurement 
$\left| \Delta e \left / e \right.\right|$ on  the ellipticity  ${e}$. The calculations have been done under the assumption that 
the azimuths $\psi_ 1$ and  $\psi_2$
can be measured with the accuracy of $0.05 ^\circ$.
It follows from Fig.\ref{pic1} that the use of  compensators with small phase retardations makes it possible to increase the accuracy and sensitivity of  ellipticity measurement.

\section{ Experimental implementation of the proposed method}


In this section, we will show how it is possible to improve the quality of the laser beam liner polarization and to demonstrate the method implementation.

The ellipticity of the radiation of a He-Ne laser is usually   $ \sim 5 \times 10^{ - 2}$. 
 It is possible to reduce the ellipticity of the laser beam up to $ \sim 10^{ - 3}$  using a  Glan prism polarize.
We improved the quality  of the laser beam linear polarization using a phase compensator (pre-compensator) with  phase retardation $ \Gamma_{\rm pc}\sim 1^\circ $. 
  
\begin{figure}[h]
\begin{center}
\includegraphics[scale=0.4]{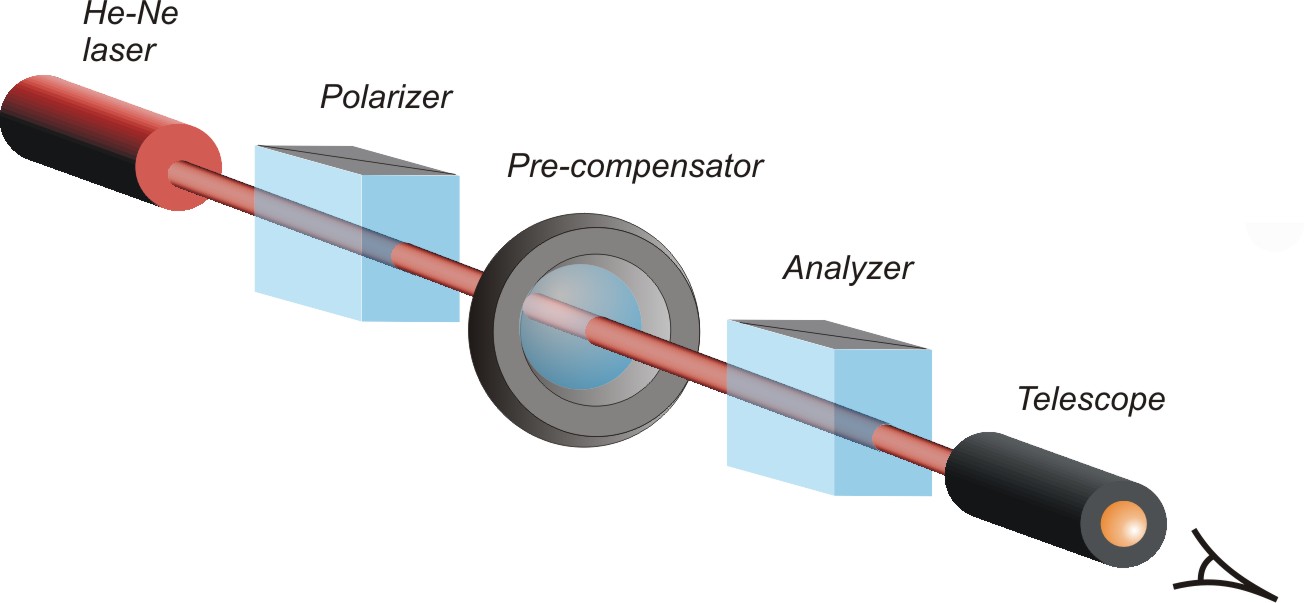}

\caption{ Experimental setup for the improvement of the laser beam linear polarization.}
\label{pic51}
\end{center}
\end{figure}

   The experimental set up for the improvement of the laser beam linear polarization is shown  in Fig. \ref{pic51}.
The   experimental setup consisted of an He-Ne laser operating at wavelength $\lambda= 0.63$ $\rm \mu m$, two Glan prisms served as a polariser and an analyser, a pre-compensator and a telescope.

We crossed the polariser and analyser to improve laser beam linear polarization. The pre-compensator was between the crossed polariser and analyser and we rotated the pre-compensator until we observed the total light extinction.
The telescope, focused
at infinity, allowed us to determine the position of the maximum darkness with  higher accuracy. 

\begin{figure}[h]
\begin{center}
\includegraphics[scale=0.4]{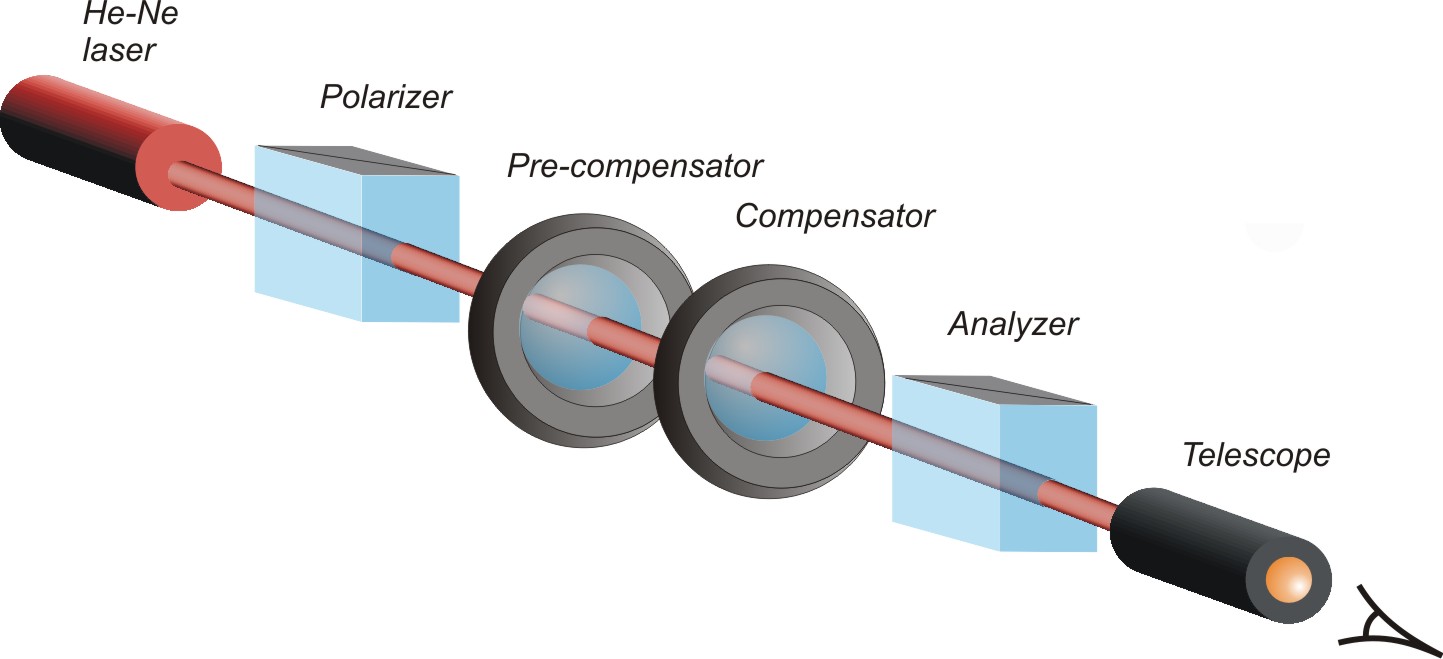}
\caption{ Experimental setup for the improved linear polarized  laser beam ellipticity measurement.}
\label{setup2}
\end{center}
\end{figure}

We measured the obtained laser beam ellipticity  using the proposed method.
We chose a compensator with the phase retardation $\Gamma \sim 1^\circ$ for the implementation of this method. This  compensator would allow us to measure ellipticities up to 
$\sim 8\times 10^{-3}$ 
(Fig. \ref{pic2}).
The compensator consisted of  two phase plates with a small difference in their phase retardations.
 The phase plates were placed parallel to each other in such a way that the slow axis of one plate was parallel to the fast axis of the second plate. The resulting phase shift of the compensator was equal to $\Gamma =0.96 \pm 0.01^\circ$. 
The phase retardation was measured using the methods described in \cite{Bibikova2006,Kundikova2010a}.
The compensator was put between the pre-compensator and  the analyser (Fig.\ref{setup2}).
To measure the obtained laser beam ellipticity we rotated the compensator along with the analyse rotation and determined  two angular  positions of the compensator $\psi_1$ and $\psi_2$, which corresponded to light extinction
 and determined the sign of $\beta$. Using Eqs.(\ref{eq05},\ref{eq5}) we calculated the value of the   ellipticity angle $\varepsilon$. It should be stressed that  the ellipticity $e$ and   ellipticity angle $\varepsilon$ are equal ($e \approx \varepsilon  $) in the range of $ 1.0 \times 10^{-4}  \div  8.0\times 10^{-3}$.

\begin{figure}[h]
\begin{center}
\includegraphics[scale=0.5]{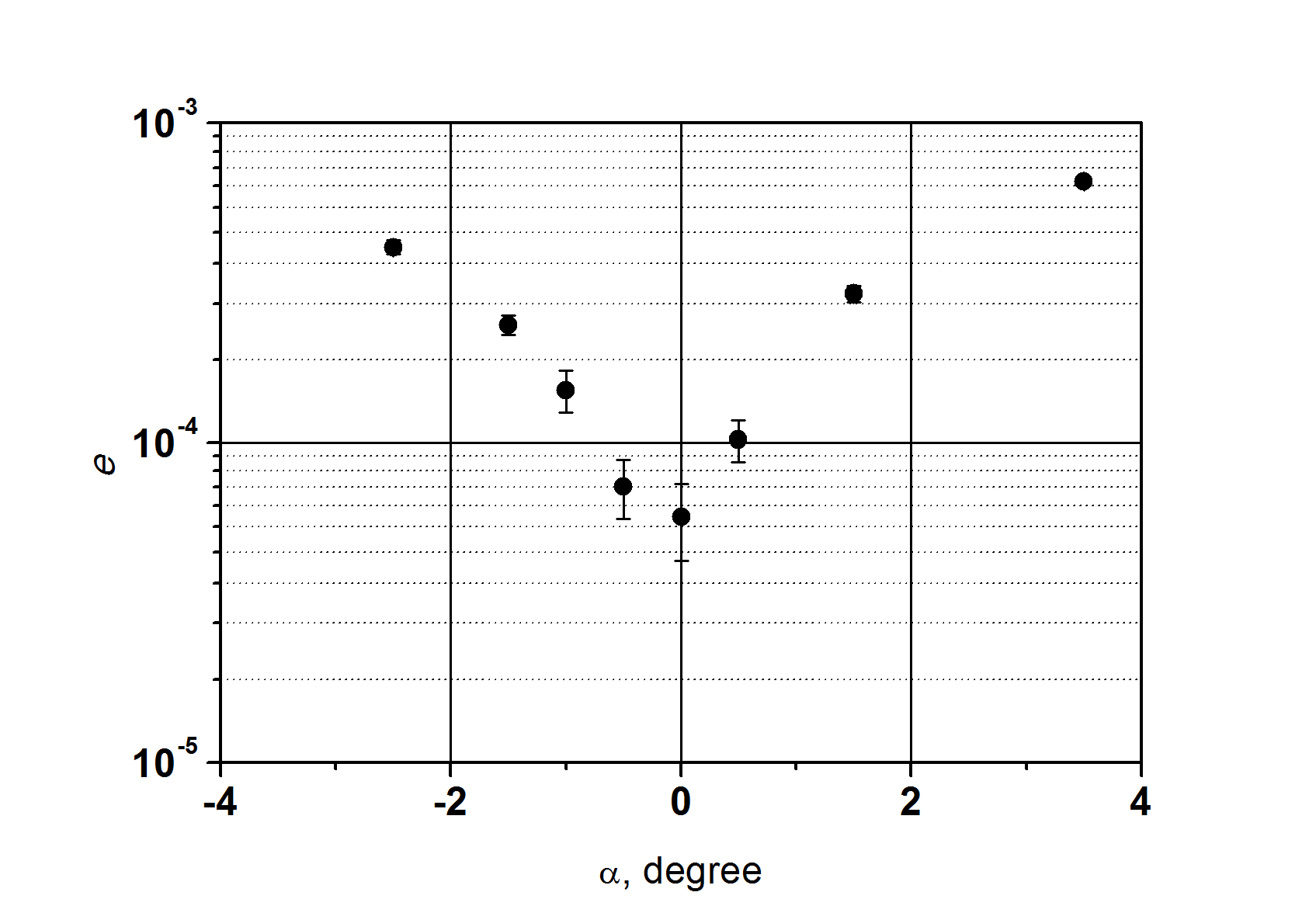}
\caption{ Dependence of the ellipticity of the light beam $e$ on pre-compensator azimuth $\alpha $ (angular position).}
\label{pic3}
\end{center}
\end{figure}

To prove that the obtained ellipticity is the best ellipticity and to test the proposed method, we rotated the pre-compensator in the range of  pre-compensator azimuth  $-2.5^\circ \leq \alpha \leq 3.5^\circ$ around the compensator angular position $\alpha_0$ corresponding to the total extinction and we measured the resulting ellipticities of the light beam.
The obtained values of ellipticities are shown in Fig. \ref{pic3}. 
Error bars were calculated according to the data presented in Fig. \ref{pic1}. 
One can see from Fig. \ref{pic3} that the ellipticity of the light beam corresponding to the position of the pre-compensator total extinction $\alpha_0$ is equal to $5 \times10^{-5}$. Slight pre-compensator rotation in the vicinity of the null point $\alpha_0$ leads to the ellipticity increase up to  $5\times10^{-4}$. 

After the measurements we installed the pre-compensator in the null position $\alpha_0$ and considered the obtained light beam with ellipticity $e\approx 5\times 10^{-5}$ as a linear polarized beam.

\section{Estimation of the accuracy of the method} 
We need to generate elliptically polarized light with the known ellipticity to estimate the accuracy of the method. It is possible to set different values of the light beam   ellipticities by the conversion of linear polarized light into elliptically polarized light using a  phase plate with the known  phase retardation $\Gamma_{\rm s}$ and amplitude relative absorption $F_{\rm s}$. The dependence of ellipticity on the phase retardation $\Gamma_{\rm s}$, amplitude relative absorption $F_{\rm s}$, and  azimuth of the input linear polarization $\beta_{ \rm s}$ is given by \cite{Bibikova2006}: 
\begin{equation}
\sin2\varepsilon_{\rm s}= \frac{F_{ \rm s}\tan\beta_{ \rm s}\sin\Gamma_{\rm s}}{1+F_{ \rm s}^2\tan\beta^2_{ \rm s}}.
\label{beta}
\end{equation}
Varying the azimuth of the linearly polarized light  at the input of this setting retardation plate $\beta_{ \rm s}$,
we can receive elliptically polarized light with the required  ellipticity. 
If we are going to work in the range of ellipticities up to $\sim 8\times 10^{-3}$ we should take into account that $\sin 2 \varepsilon\approx 2\varepsilon$ and $e \approx \varepsilon$. It follows from Eq. (\ref{beta}) that if we use a quarter wave plate as the setting retardation plate we should choose  the angle $\beta_{ \rm s} < 0.5^\circ $. According to Eq. (\ref{beta}), decrease of $\Gamma_{ \rm s}$ leads to increase of  $\beta_{ \rm s}$ values.

We used the setting mica wave plate with the phase retardation 
 $\Gamma_{\rm s}  =  21.60\pm 0.05^\circ $ and relative amplitude transmission coefficient $ F_{\rm s}=0.971\pm 0.001$ to increase values of $\beta_{ \rm s}$ up to $\sim 1^\circ$. The setting wave plate parameters were  determined by the methods described in Ref.\cite{Bibikova2006,Kundikova2010a}. 
 
\begin{figure}[h]
\begin{center}
\includegraphics[scale=0.4]{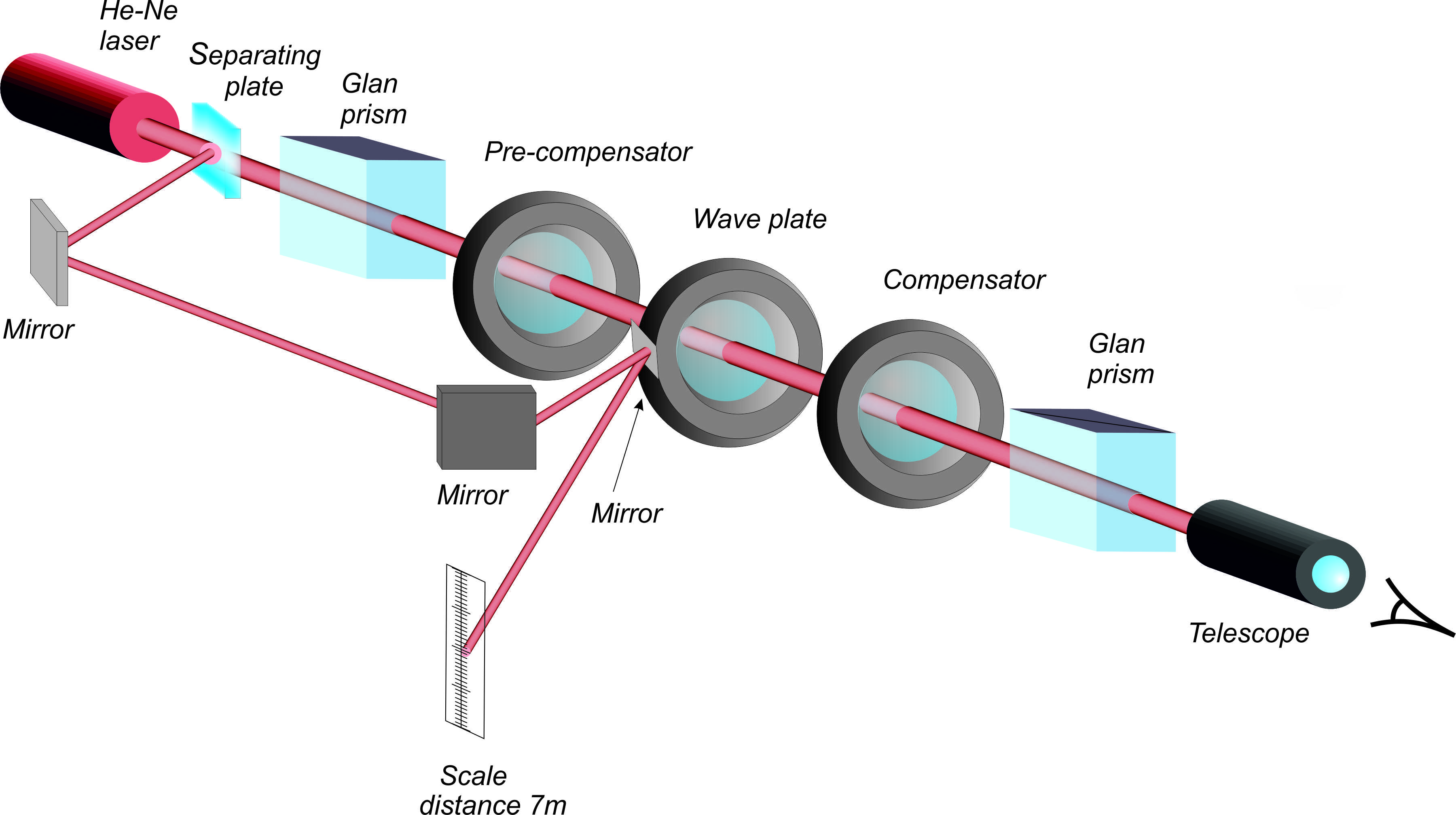}
\caption{ Experimental setup for the  method accuracy estimation.}
\label{pic5}
\end{center}
\end{figure}

 
The experimental setup shown in Fig.\ref{pic5}
 consisted of an He-Ne laser operating at wavelength $\lambda= 0.63$ $\rm \mu m$, a semitransparent plate, two mirrors, a scale placed at the distance of 7 m from the experimental setup, the polariser and analyser, the pre-compensator, the setting wave plate, the compensator, and the telescope. As mentioned above, we used the polariser and the pre-compensator to obtain a laser beam with ellipticity $e\approx 5\times 10^{-5}$ and considered this beam  as the linear polarized beam. The compensator, analyser, and telescope were used for the ellipticity measurement. The setting mica wave plate was used to set the required elliptical polarization of the beam.
 
 To set light ellipticity with high accuracy we should set the position of the setting plate with high accuracy  
 using a special technique. A small part of the laser radiation was reflected by the semitransparent plate and  was directed using two  mirrors to a small mirror fixed at  the rim of the setting  wave plate. The light beam reflected from the setting plate mirror was fixed at  a millimetre scale, located at a distance of 7 meters from the rotation axis of  the setting wave plate. The position of the reflected beam made it possible to calculate the azimuth of the input polarization $\beta_{\rm s}$ with an accuracy not worse than $0.005^
\circ$.
 
\begin{figure}[h]
\begin{center}
\includegraphics[scale=0.5]{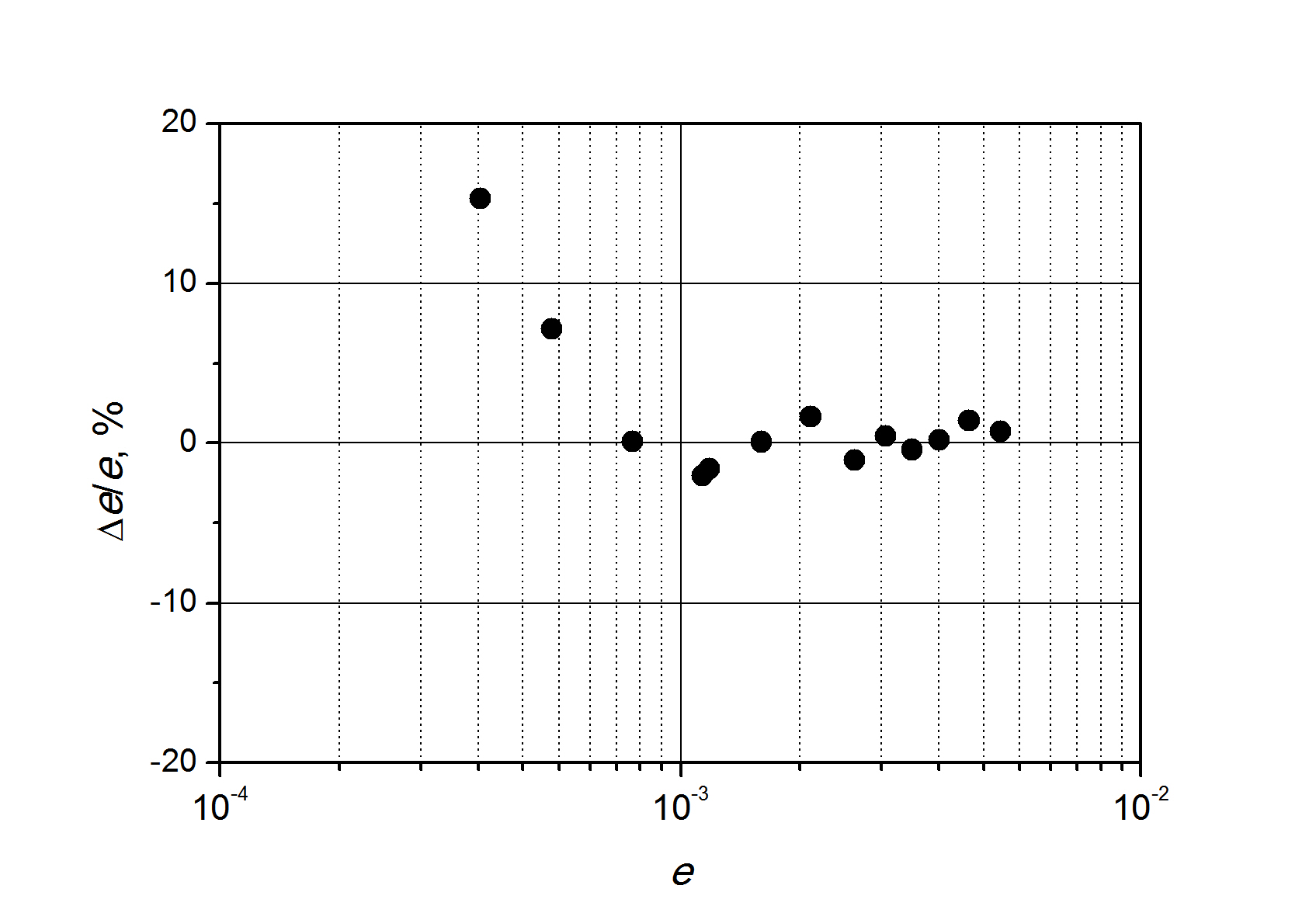}
\caption{ Relative error on the determination of the ellipticity 
$\Delta e
\left/e
  \right.$ as a function of  ellipticity $e$.
}
\label{pic6}
\end{center}
\end{figure}

 We set angle $\beta_{\rm s}$ and calculated ellipticity $e_{\rm s}$ to set the required ellipticity. After setting the known ellipticity we used the experimental procedure described in section 3 to measure the setting ellipticity. The compensator, analyser, and  telescope were used to make the measurement. We determined the difference between the setting value of ellipticity $e_{\rm s}$ and the experimentally obtained value of ellipticity $e_{\rm exp}$ to calculate the method accuracy. The values of $(e_{\rm s}-e_{\rm exp})/e_{\rm s}=\Delta e/e$ as a function of the ellipticity itself $e=e_{\rm s}$ are shown in Fig. \ref{pic6}.
  One can see from   Fig. \ref{pic6} that relative error of the ellipticity determination  $\Delta e/e$ does not exceed $ 2\%$ in the range 
 $7.5\times 10^{ - 4} \div 4.5\times 10^{ -3}$. Nevertheless, the error greatly increases in the interval of the ellipticities $3.0 \times 10^{ -4} \div 7.0\times 10^{ -4}$. The value of the error is connected with the setting ellipticity uncertainty for ellipticities less than $7.0\times 10^{ -4}$. That is why we carried out the measurement in the range of $3.0 \times 10^{ - 4} \div 4.5\times 10^{ -3}$.
 


\section*{Conclusion}
We propose a precise ellipsometric method for the investigation of coherent light in the wide range of polarization ellipticities including small ellipticities. The main feature of this method is the use of a compensator with the phase delay providing the maximum accuracy of measurements for each selected range of ellipticities and taking into account the interference of multiple reflections of coherent light. 
 The experimental measurements were implemented in the range of   $3.0 \times 10^{ -4 } \div 4.5\times 10^{ -3}$. The relative error of the ellipticity measurement in the range  of $7.5\times 10^{ - 4} \div 4.5\times 10^{ -3}$ does not exceed $ 2\%$.


\section*{Acknowledgement}
This work was partly carried out within the scope of the topic of State Assignment No. 0389-2014-0030.


\end{document}